\documentclass{llncs}

\usepackage{paradise}
\usepackage{longtable,booktabs}
\usepackage{graphicx}
\usepackage[sectionbib]{natbib}
\setcitestyle{numbers,square,comma}

\newcommand{\hypertarget}[2]{#2}

\usepackage{caption}
\begin{document}
\bibliographystyle{plainnat}

\pagestyle{plain}
\mainmatter

\title{A Simulator for \llvm{} Bitcode\thanks{This work has been partially supported by the Czech Science Foundation grant No.~18-02177S.}}

\author{Petr Ročkai \and Jiří Barnat}
\institute{\fimuni\\ \{xrockai,barnat\}@fi.muni.cz}

\maketitle

\begin{abstract}
  In this paper, we introduce an interactive simulator for programs in the form of \llvm{} bitcode. The main features of the simulator include precise control over thread scheduling, automatic checkpoints
  and reverse stepping, support for source-level information about functions and variables in C and C++ programs and structured heap visualisation. Additionally, the simulator is compatible with \divm{}
  (\divine{} VM) hypercalls, which makes it possible to load, simulate and analyse counterexamples from an existing model checker.
\end{abstract}

\hypertarget{introduction}{%
\section{Introduction}\label{introduction}}

Verification tools are increasingly adopting \llvm{} bitcode as their input language of choice. A frequent reason for implementing \llvm{}-based model checkers (and other analysis tools) is that they can
leverage existing compiler front ends, CLang in particular. This in turn enables those model checkers to work with C and even C++ programs without dealing with their irregularity and complexity.
Clearly, this tremendously improves the usefulness of any such tool, since C and C++ are widespread implementation languages, and implementation-level model checking is naturally desirable for many
reasons.

An additional benefit of the standardisation around the \llvm{} IR~\citep{llvm16:llvm.languag} (intermediate representation) is that an ecosystem of tools is emerging, where those tools can cooperate
through the common input format. Analysis and model checking tools can be used to ascertain correctness of the program with respect to a specification; however, when they find that there is a
violation, printing ``property violated'' is rarely enough. For the result to be genuinely useful, it must somehow convey \emph{how} the specification is violated to the user, so they can analyse the
problem and fix their program. One option is to print a \emph{counterexample trace}, which describes the violating execution of the program. In traditional model checkers, for example, it is often
sufficient to provide a textual description of the entire execution, since the input model is usually small and its states and transitions can be described compactly.

More advanced tools, however, provide a \emph{simulator}, an interactive tool for stepping through the counterexample, where the user can highlight and investigate particular sections of the
counterexample in more detail, and fast-forward through other, uninteresting parts. A simulator is often also useful as an exploratory tool: the behaviour of the system can be explored by the user,
manually navigating through its state space and inspecting variables along the way.

In case of C and C++ programs, it is vitally important that counterexamples can be inspected interactively, since the state of a program is a very complicated structure, often comprising hundreds of
kilobytes of structured data. Moreover, violating executions can be quite long, easily hundreds or thousands of distinct states, with non-trivial relationships.

The main contribution of this paper is a reusable simulator for C and C++ code. Since it builds on the \llvm{} intermediate language, it can be used by multiple different tools which produce
counterexamples or otherwise work with \llvm{} bitcode, and is easily adapted to new high-level languages with \llvm{} toolchains (like Objective C or Rust). To the best of our knowledge, this work is
unique in the sense that no other simulator which would handle C++ programs is available, and simulators which handle C code typically miss important features. Moreover, the simulator is also
reusable: while originating from the \divine{} tool set, it can be used standalone, or possibly in combination with other analysis and verification tools.

From a more theoretical standpoint, the \emph{debug graph} (described in Section~\ref{sec:debugnode}) represents a new approach to reconciling low-level data as it exists during program runtime with
the high-level structure declared in the source code. Another new idea is to build a simulator based on compiled code (as opposed to interpreting the source code directly) and leveraging existing
debugger-focused infrastructure (debug metadata in particular), making the implementation especially simple and compact.

The rest of this paper is structured as follows: in Section~\ref{sec:related} we discuss related work and compare our approach to existing tools. Section~\ref{sec:llvm} describes the \llvm{} bitcode as
it is used by the simulator, how the simulator represents the program state and also introduces the \emph{debug graph}. The focus of Section~\ref{sec:data} is presentation of the data aspects of a
program, while Section~\ref{sec:statespace} is concerned with the program's state space. Section~\ref{sec:implementation} mentions some of the more important implementation details.
Section~\ref{sec:conclusion} wraps the paper up. Additional resources (mainly evaluation-related) are available online\footnote{\url{https://divine.fi.muni.cz/2019/sim/}}.

\hypertarget{sec:related}{%
\section{Related Work}\label{sec:related}}

It is a well-established fact that isolating some bad behaviour of a program in a test is, in itself, not sufficient to easily explain the cause of the problem~\citep{ball03:from}. The situation is
similar in (linear-time) model checking, where a counterexample trace can often be extracted easily enough, but it may not contain sufficient detail, or conversely, may swamp the user in large amount
of irrelevant data~\citep{visser02:what.went.wrong}. The problem also goes beyond the software realm, as witnessed in, for instance, verification of MATLAB Simulink
designs~\citep{barnat12:tool.chain}.

There are basically two orthogonal approaches that attempt to resolve these problems. One is to locate, or at least narrow down, the error automatically, in the hopes that from such a narrowed-down
trace, the user will be able to understand the problem by inspection of the source code. In the domain of software verification, this approach is pursued by many tools: counterexamples for violation
of temporal properties, generated by the software model checker SLAM~\citep{ball04:slam.static}, for instance, can be analysed and reduced to only cover a small number of source lines, in which the
root cause of the error is most likely to lie~\citep{ball03:from}. An approach to succinctly describe assertion violations (violations of safety properties), based on automated dependency analysis,
has also been proposed~\citep{basu12:gettin.root.problem}. Finally, counterexamples from CBMC can be post-processed, in an approach similar to those mentioned above, with a tool called
\texttt{explain}~\citep{groce04:unders.counter}, in this case based on distance metrics.

Unfortunately, even if the problem area is only a few lines of source code, it can be very hard to understand the dynamic behaviour during the erroneous execution. The problem gets much worse when the
program in question is parallel, because reasoning about the behaviour of such programs is much harder than it is in the sequential case.

To make understanding and fixing problems in programs (or complex systems in general) easier, many formal verification tools come equipped with a simulator. For instance the UPPAAL tool for analysis
of real-time systems provides an integrated graphical simulator~\citep{behrmann04:tutorial.uppaal}. Another example of a formal analysis tool with a graphical simulator would be
LTSA~\citep{magee99:behavior.analys}, based on labelled transition systems as its modelling formalism.

Like many verification tools, the \texttt{valgrind}~\citep{nethercote07:valgrin} run-time program analyser is primarily non-interactive, but it provides an interface to allow interactive exploration
of program state upon encountering a problem, based on \texttt{gdb}~\citep{stallman10:debugg}.

Our simulator is based on \divm{}~\citep{rockai18:divm}, an extension of the \llvm{} language that allows verification and analysis of a wider class of programs (a more detailed description of the \divm{}
extensions is given in Section~\ref{sec:extensions}). Since pure \llvm{} is retained as a subset of the \divm{} language, the simulator can also transparently work with pure \llvm{} bitcode.

Besides its relationship to various simulators for modelling and design languages, a simulator for \llvm{} bitcode is, through its application to code written in standard programming languages like C,
related to standard symbolic debuggers. A ubiquitous example on \posix{} systems is \texttt{gdb}, the GNU debugger~\citep{stallman10:debugg}. Unlike a simulator, which interprets the program, a debugger
instead attaches to a standard process executing in its native environment. A more recent example would be \texttt{lldb}~\citep{lee13:lldb}, which works in essentially the same way, but builds on \llvm{}
components.

\hypertarget{sec:debuggers}{%
\subsection{Comparison to Symbolic Debuggers}\label{sec:debuggers}}

As outlined above, simulators and debuggers substantially differ in their mode of operation and this leads to very different overall trade-offs. For example, a simulator is much more resilient to
memory corruption than a debugger, because the latter has only limited control over the process it is attached to. Both types of tools rely on understanding the execution stack of the program;
however, if the program corrupts its execution stack, a debugger must rely on imprecise heuristics to detect this fact and risks providing wrong and possibly misleading information to the user. The
simulator can, on the other hand, quite easily prevent such corruption from happening, since it simulates the program at instruction level, and can enforce much stricter memory protections.

On the other hand, the situation is reversed when the program interacts with its surroundings through the operating system. In a debugger, such communication comes about transparently from the fact
that the program is a standard process in the operating system and has all the standard facilities at its disposal. In a simulator, communication with the operating system must be specifically relayed
and due to imperfections in this translation, some programs may misbehave in the simulation.

Finally, a simulator has a substantial advantage in two additional areas: first, a simulator can very precisely and comfortably control thread interleaving. This allows analysis of subtle
timing-dependent issues in the program. Second, since a simulator has a complete representation of the program's state under its control, it can easily move backwards in time or compare variable
values from different points in the execution history. While both scheduler locking and reversible debugging exist to a certain degree in traditional debuggers~\citep{visan11:urdb}, those features are
very hard to implement and usually quite limited.

\hypertarget{sec:llvm}{%
\section{\llvm{} Bitcode}\label{sec:llvm}}

The \llvm{} bitcode (or intermediate representation)~\citep{llvm16:llvm.languag} is an assembly-like language primarily aimed at optimisation and analysis. The idea is that \llvm{}-based analysis and
optimisation code can be shared by many different compilers: a compiler front end builds simple \llvm{} IR corresponding to its input and delegates all further optimisation and native code generation to
a common back end. This architecture is quite common in other compilers: as an example, GCC contains a number of different front ends that share infrastructure and code generation. The major
innovation of \llvm{} is that the language on which all the common middle and back end code operates is exposed and available to 3rd-party tools. It is also quite well documented and \llvm{} provides
stand-alone tools to work with both bitcode and textual form of this intermediate representation.

From a language viewpoint, \llvm{} IR is in a partial SSA form (single static assignment) with explicit basic blocks. Each basic block is made up of instructions, the last of which is a
\emph{terminator}. The terminator instruction encodes relationships between basic blocks, which form an explicit control flow graph. An example of a terminator instruction would be a conditional or an
unconditional branch or a \texttt{ret}. Such instructions either transfer control to another basic block of the same function or stop execution of the function altogether.

Besides explicit control flow, \llvm{} also strives to make much of the data flow explicit, taking advantage of partial SSA for this reason. It is, in general, impossible to convert entire programs to a
full SSA form; however, especially within a single function, it is possible to convert a significant portion of the code. The SSA-form values are called \emph{registers} in \llvm{} and only a few
instructions can ``lift'' values from memory into registers and put them back again (most importantly \texttt{load} and \texttt{store}, respectively, plus a handful of atomic memory access
instructions).

From the point of view of a simulator, memory and registers are somewhat distinct entities, both of which can hold values. Memory is completely unstructured at the \llvm{} level, the only assumption is
that it is byte-addressed (endianity of multi-byte values is configurable, but uniform). Traditional C stack is, however, not required. Instead, all ``local'' memory is obtained via a special
instruction, \texttt{alloca}, and treated like any other memory (memory obtained by \texttt{alloca} is assumed to be freed automatically when the function that requested the memory exits, via
\texttt{ret} or any other way, e.g.~due to stack unwinding during an exception propagation). Therefore, a C-style stack is a legitimate way to implement \texttt{alloca}, but not the most convenient in
a simulator (for more details on how memory is handled in our simulator, see Section~\ref{sec:memory}).

\hypertarget{sec:extensions}{%
\subsection{Verification Extensions}\label{sec:extensions}}

Unfortunately, \llvm{} bitcode alone is not sufficiently expressive to describe real programs: most importantly, it is not possible to encode interaction with the operating system into \llvm{} instructions.
When \llvm{} is used as an intermediate step in a compiler, the lowest level of the user side of the system call mechanism is usually provided as an external, platform-specific function with a standard C
calling convention. This function is usually implemented in the platform's assembly language. The system call interface, in turn, serves as a gateway between the program and the operating system,
unlocking OS-specific functionality to the program. An important point is that the gateway function itself cannot be implemented in portable \llvm{}. Moreover, while large portions of the kernel are
often implemented in C or a similar portable language, they are also tightly coupled to the underlying hardware platform.

The language of ``real'' programs is, therefore, \llvm{} enriched with system calls, which are provided by the operating system kernel. For verification purposes, however, this language is quite
unsuitable: the list of system calls is long (well over 100 functions on many systems) and exposes implementation details of the particular kernel. Moreover, re-implementing a complete operating
system inside every \llvm{} analysis tool is wasteful. To reduce this problem, a much smaller set of requisite primitives was proposed in \citep{rockai18:divm} (henceforth, we will refer to this enriched
language as \divm{}). Since for model checking and simulation purposes, the program needs to be isolated from the outside world, we can skip most of the complexity of an operating system kernel --
communication with hardware in particular. Therefore, it is possible to implement a small, isolated operating system in the \divm{} language alone. One such operating system is \dios{} -- the core OS is
about 1500 lines of C++, with additional 5000 lines of code providing \posix{}-compatible file system and socket interfaces.

Thanks to its support for the \divm{} language, our simulator can transparently load programs which are linked to \dios{} and its \texttt{libc} implementation. Since a program compiled into the \divm{}
language is fully isolated from any environment effects, it can be simulated just like a pure \llvm{} program could be.

Finally, while the simulator uses \divm{} to evaluate program instructions and hence relies on correctness of the implementation, errors in \dios{} have a smaller impact. The \dios{} code is executed in the
virtual machine, and is subject to its error checking: therefore, in this case, the most likely outcome is by far a spurious error which can be analysed using the simulator itself.

\hypertarget{sec:memory}{%
\subsection{Program Memory}\label{sec:memory}}

Internally, the simulator uses \divm{} to evaluate \llvm{} bitcode, and therefore, how memory is represented in the simulator is directly inherited from \divm{}. This means that we can take advantage of the
fact that \divm{} tracks each object stored in memory separately, and also keeps track of relationships (pointers) between such objects.\footnote{How this is achieved is described in more detail
  in~\citep{rockai18:divm}.} This way, the simulator precisely knows which words stored in memory are pointers and the exact bounds of each object in memory.

Moreover, \divm{} can efficiently store multiple snapshots of the entire address space of the program, both in terms of space (most of the actual storage is shared between such snapshots) and time
(taking a snapshot needs time roughly proportional to the total size of modified objects since the last snapshot). Once a snapshot is taken, it is preserved unmodified, regardless of the future
behaviour of the program (that is, it becomes persistent).

\begin{figure}
\hypertarget{fig:stack}{%
\centering
\includegraphics{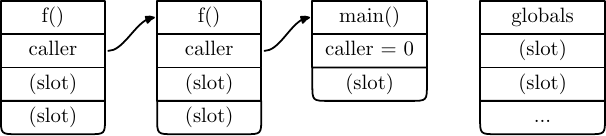}
\caption{Execution stack and global variables.}\label{fig:stack}
}
\end{figure}

The execution stack of an \llvm{} program consists of activation frames, one for each active procedure call. In \divm{}, activation frames are separate memory objects. Moreover, each memory-stored local
variable (i.e.~those represented by \texttt{alloca} instructions) is again represented by a distinct memory object. Each frame object contains 2 pointers in its header (one points at the currently
executing instruction, the other to the parent frame). Besides the header, the rest of the object is split into \emph{slots}, where each slot corresponds to a single \llvm{} \emph{register}. An example
stack structure is shown in Figure~\ref{fig:stack}. The correspondence between slots and \llvm{} registers is maintained by \divm{} and is available to the simulator.

Together, those features of \divm{} make it very easy to access the program state in a highly structured fashion. When compared to a traditional debugger, which must work with nearly unstructured memory
space, the information our simulator can provide to the user is simultaneously easier to obtain and more detailed and reliable. Finally, since \divm{} strictly enforces object boundaries, both the
control stack and heap structure in our simulator are very well protected from overflows and other memory corruption bugs in the program. Therefore, the simulated program cannot accidentally destroy
information which is vital for the functioning of the simulator, like all too often happens in debuggers.

\hypertarget{sec:debuginfo}{%
\subsection{Relating Bitcode to Source Code}\label{sec:debuginfo}}

In native code debuggers, the relationship between the binary and the original source code is often not quite obvious. For this reason, in addition to the executable binary, the compiler emits
metadata which describe these relationships. For instance, it attaches a source code location (filename and line number) to each machine instruction. This way, when the debugger executes an
instruction, it can display the relevant piece of source code. Likewise, it can analyse the execution stack to discover how the currently executing function was called, and display a \emph{backtrace}
consisting not only of function names, but also source code lines. This is important whenever a given function contains two similar calls.

\begin{figure}
\hypertarget{fig:struct}{%
\centering
\includegraphics{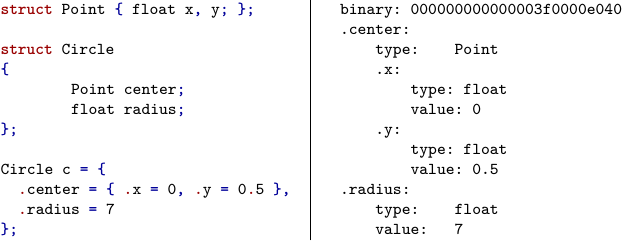}
\caption{An example C \texttt{struct} type and the corresponding representations: binary and structured (the latter is only possible with debug metadata).}\label{fig:struct}
}
\end{figure}

The situation is analogous in \llvm{}-based tools. Compiler front ends are therefore encouraged to generate \emph{debuginfo metadata} (in a form that reflects the structure of the DWARF debug information
format, which is widely used by native source-level debuggers). Besides the vitally important source code locations, the metadata describe local and global variables and their types (including user
defined types, like \texttt{struct} and \texttt{union} types in C). This in turn enables the debugger to display the data in a structured way, resembling the structure which exists in the source code.
For example, \texttt{struct} types in C have named fields -- the debugger can use the debug metadata to discover the relationship between offsets in the binary representation of the value with the
source-level field names (an example is shown in Figure~\ref{fig:struct}).

\hypertarget{sec:debugnode}{%
\subsection{Debug Graph}\label{sec:debugnode}}

\begin{figure}
\hypertarget{fig:heap}{%
\centering
\includegraphics[width=1\textwidth,height=0.4\textheight]{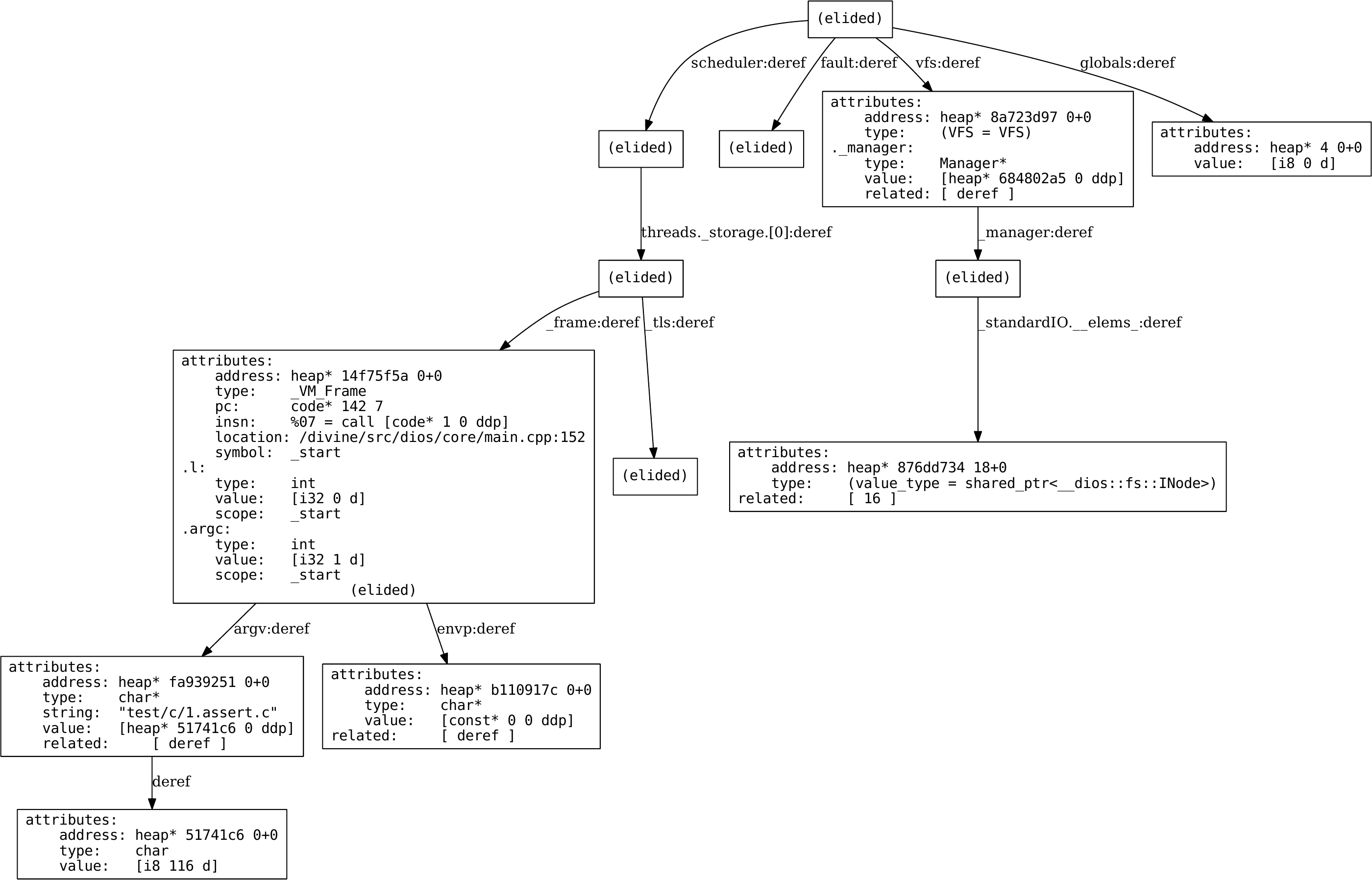}
\caption{A debug graph of a simple program. A single memory object may contain multiple \emph{component} debug nodes which are rendered textually. The arrows correspond to \emph{relations}. The
depicted graph was obtained directly from the simulator; the only change was that descriptions of some of the nodes were elided for presentation purposes.}\label{fig:heap}
}
\end{figure}

The memory graph maintained by \divm{} is a good basis for presenting the program state to the user, but on its own is insufficient: the only type information it contains is whether a particular piece of
memory holds a pointer or not. Therefore, we overlay another graph structure on top of the memory (heap) graph, with richer type information based on debuginfo metadata (more details on how this graph
is computed will be presented in Section~\ref{sec:data}). The nodes in the debug graph may be further structured: they have \emph{attributes} (atomic properties, such as an integer or floating point
value), \emph{components} and \emph{relations}. While both components and relations are again nodes of the graph, they crucially differ in how they relate to the underlying memory: components of a
debug node represent the same memory as their parent node; for example, a debug node which consists of a \texttt{struct} C type will contain a \emph{component} for each field of the \texttt{struct}.
In contrast, \emph{relations} of a debug node correspond to the pointers embedded in the memory it represents (it may, however, point back at the same object it is embedded in). An example debug graph
is shown in Figure~\ref{fig:heap}.

Since memory objects are \emph{persistent} in \divm{} (cf. Section~\ref{sec:memory}), so is the debug graph in our simulator. This means that objects (debug nodes) are immutable, i.e.~they always come
from a \emph{snapshot} of the memory of the program. Since it would be too expensive to make a copy of the entire memory after every instruction, such snapshots are implemented via copy-on-write
semantics.

\hypertarget{sec:data}{%
\section{Working with Data}\label{sec:data}}

Providing facilities for inspecting data of the program is one of the main functions of an interactive debugger or a simulator. This data can be presented in different forms and from different
starting points. In our simulator, heap memory is structured explicitly as a graph, and we can leverage this to greatly improve presentation of data. An example of such a graph is shown in
Figure~\ref{fig:heap}. Each node of the graph corresponds to a single in-memory object, which can have (and often has) additional internal structure. The internal structure reflects the C/C++ type
which is deduced from the types of pointers pointing at this particular node.

\hypertarget{starting-points}{%
\subsection{Starting Points}\label{starting-points}}

For certain memory objects, the type information is directly encoded in the metadata generated by the compiler and does not need to be inferred via pointers. Such objects are the starting point of the
type assignment process by which the \emph{debug graph} is obtained.

In principle, there are 2 types of such objects: \emph{activation frames} and \emph{globals}. Both consist of \emph{slots} which in turn contain values. Values in those slots either correspond to
values of (local or global) source-level variables, or contain pointers to variables held elsewhere in memory. In both cases, a \emph{component} debug node is created for each slot, based on the debug
information generated by the compiler. These components then form a basis for presenting the data to the user.

Additionally, in \divm{}, there is always a single distinguished \emph{root object} in the heap, from which the entire heap is reachable, including the stacks of all threads and any kernel data
structures. The address and the C type of this \emph{root object} is also available to the simulator, and is mainly used to discover all the nodes of the 2 abovementioned types.

\hypertarget{typing-the-heap}{%
\subsection{Typing the Heap}\label{typing-the-heap}}

In all cases, the type information available for the starting points is used to derive type information for the portion of the heap reachable from that starting point. For frames, we can deduce which
function the frame belongs to, and obtain information about the frame layout used by that function. That is, for each \llvm{} register, we obtain a corresponding C type, which is usually either a
primitive type or a pointer. If the type is a pointer and it is not null or otherwise invalid, there is an edge in the graph of the heap corresponding to this pointer. The object at the other end of
the edge is then assigned the \emph{base type} of the pointer, that is, type of a value obtained by dereferencing the pointer. This procedure is then repeated recursively until all objects where type
information exists are assigned a type.

Of course, there is a potential for ambiguity: not all C/C++ programs are consistently typed, therefore, multiple edges pointing at a single object can each carry a different type. In this case, we
collect all the applicable types and construct a synthetic \emph{union type}, which is assigned to any such ambiguous debug node. This ambiguity might propagate downstream from an affected node, but
for most programs, this does not appear to pose a significant problem.

\hypertarget{sec:data_and_control}{%
\subsection{Relating Data and Control}\label{sec:data_and_control}}

The control flow of a C program is reflected in the execution stack and is a part of the program's data. C and C++ are lexically scoped languages: which variables are currently in scope depends on
which function (and possibly which block in that function) is currently executing. This is achieved by making local variables part of the execution stack: when a function is entered, an
\emph{activation frame} (or activation \emph{record}) is pushed onto the execution stack. In a native execution environment, the frame has space for CPU register spills and for local variables which
have their address taken. In \divm{}, there are no general-purpose registers as such; instead, \llvm{} registers are stored inside the frame itself. Any address-taken variables are stored as separate
objects (and their address is stored in a register).

Additionally, in a typical implementation of C, the activation frame contains a \emph{return address}, which is a pointer to the \texttt{call} instruction that caused the current function to execute.
In \divm{}, the frame instead contains a \emph{program counter} (in a real CPU, the program counter, also known as instruction pointer, is held in a register). The program counter tells us which
function, and which instruction within that function, is currently being executed. Each instruction can in turn be tied, via debug metadata (cf. Section~\ref{sec:debuginfo}), to a particular source
code location (a source file and a line number).

\begin{figure}
\hypertarget{fig:source}{%
\centering
\includegraphics{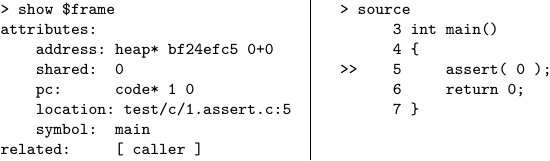}
\caption{An example interaction: listing source code.}\label{fig:source}
}
\end{figure}

As an example of how this is used in the simulator, if the user requests to list the source code of the currently executed function, the simulator examines the current active activation frame to find
the current value of the \emph{program counter}. Then it proceeds to read the corresponding debug metadata to obtain the source code file name, reads the source file, finds the line corresponding to
the program counter and prints the surrounding function (example output is shown in Figure~\ref{fig:source}).

\hypertarget{sec:statespace}{%
\section{Navigating the State Space}\label{sec:statespace}}

\begin{figure}
\hypertarget{fig:prog}{%
\centering
\includegraphics{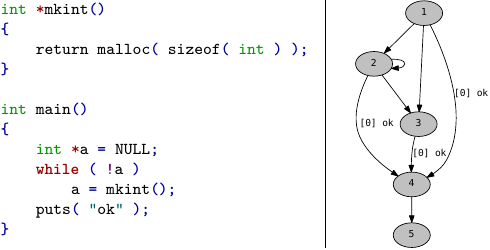}
\caption{An example C program along with its state space.}\label{fig:prog}
}
\end{figure}

If we treat the data of a program as a spatial dimension, it is natural, then, to treat the state space -- the behaviour of program as it executes -- as a time dimension. Since the state space is a
graph (cf. Figure~\ref{fig:prog}), the predecessors of a given state (the path from the initial state to the ``current'' state -- the one that is being examined) constitute the \emph{past} of the
computation. The successors, on the other hand, correspond to possible \emph{futures} of the computation (since the behaviour of the program is often non-deterministic\footnote{The behaviour of the
  program may depend on external factors, such as scheduling choices, user inputs, asynchronous events and so on.}, there is more than one possible future). In this correspondence of the state-space
graph to temporal behaviour of the program, cycles in the state space clearly correspond to behaviours that go on forever.

In a standard debugger, time can only flow in one direction, and which of the potential futures is realised can be influenced, but not controlled. In a simulator, however, it is possible to both go
backwards in time (rewind the program state to some past configuration) and to pick exactly which future should be explored. Likewise, it is entirely possible to go back in time and select a different
future to explore. These capabilities are derived mainly from the persistent and compact memory representation (see Section~\ref{sec:memory}).

\hypertarget{stepping-forward}{%
\subsection{Stepping Forward}\label{stepping-forward}}

On the other hand, the state space as explored by model checkers is often too coarse to follow the computation in detail. The states typically correspond to locations where threads interleave or where
cycles can potentially form. At this level, the edges in the state space correspond, approximately, to atomic actions in the program. Even in heavily parallel programs, though, such atomic actions
will span many instructions and possibly multiple source lines. A simulator which works at this level\footnote{This is often the case in verification-centric tools, partly because it is a simple
  implementation strategy that builds on the same primitives as the verification tool itself.} can only present very coarse computation steps to the user and not seeing the intermediate state of the
program can prevent users from relating effects to their causes. If the simulator operates with fixed computation steps, the opposite problem can also happen: the user must step through a large number
of irrelevant program configurations~\citep{kleiman93:tales}, again frustrating the debugging effort.

In contrast, debuggers give the user very precise control over the forward execution of the program, down to stepping one machine instruction at a time. However, they also make it very easy to fast
forward through thousands of lines of code, stopping when a predetermined condition is met, most often a particular source code line is executed (this feature is known as a \emph{breakpoint}).

Building the simulator on top of \divm{}, however, gives us execution control at the level of individual \llvm{} instructions, analogous to a debugger. Building on the instruction stepping mechanism, the
simulator also provides all the control functionality common in debuggers: source-line stepping -- both into and over function calls -- and various breakpoint types (on a source line or a on a
function entry).

\hypertarget{sec:going_back}{%
\subsection{Going Back}\label{sec:going_back}}

In general, it is impossible to execute individual instructions backwards. However, if execution is perfectly repeatable (as it is in a simulator), we can reach any earlier configuration of the
program by replaying the current execution from the start and stopping right before the instruction of interest executes.

Additionally, the simulator stores intermediate states (automatically at convenient locations, or at a user request). It is then possible to go back to any such stored state and continue execution
from that point. This can make the abovementioned process considerably more efficient: it is enough to replay execution from the most recent stored state that lies on the current execution path.

\begin{figure}
\hypertarget{fig:backtrace}{%
\centering
\includegraphics{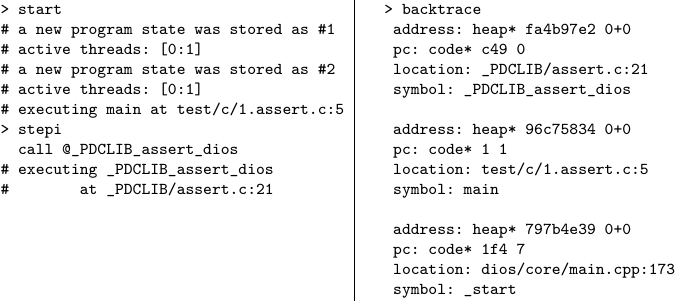}
\caption{Left: new states are discovered during execution of a program. Right: displaying a backtrace.}\label{fig:backtrace}
}
\end{figure}

\hypertarget{inspecting-the-stack}{%
\subsection{Inspecting the Stack}\label{inspecting-the-stack}}

As explained in Section~\ref{sec:data_and_control}, the control flow of a C program (or, more generally, any \llvm{} program) is tracked by a simple data structure stored in memory along with other data.
This data structure often represents the best means for a user to locate themselves within the execution of a program. A so-called \emph{backtrace} (or \emph{stack trace}) is a fundamental program
analysis tool. A backtrace lists each activation record in the (reverse) order of activation, and constitutes a description of a location in the computation of the program\footnote{This description is
  necessarily incomplete, being much more concise than the real representation of the program's state. Including additional information improves completeness, but compromises brevity, which is an
  important strength of this presentation format.} (an example is shown in Figure~\ref{fig:backtrace}).

\hypertarget{thread-interleaving}{%
\subsection{Thread Interleaving}\label{thread-interleaving}}

As mentioned in Section~\ref{sec:debuggers}, a simulator can precisely control thread interleaving: the underlying virtual machine provides means to switch threads at all relevant points. However,
many instruction interleavings have equivalent effects, and for this reason, allowing threads to be switched at arbitrary points would be wasteful. For this reason, \divm{} explicitly marks points in the
instruction stream where threads may be switched, and this behaviour is carried over to the simulator. These \emph{interrupt points} are inserted in such a manner that all possible behaviours of the
program are retained in the state space. From a simulation point of view, the downside is that the interleaving may not be the most intuitive, but the reduction in the number of possible states
generally outweighs this, since the user needs to consider fewer runs. To further reduce the number of context switches, a model checker may use some form of partial order reduction, but this is not
necessary in a simulator, since it doesn't need to explore or store the entire state space.

\hypertarget{simulating-counterexamples}{%
\subsection{Simulating Counterexamples}\label{simulating-counterexamples}}

There are two major tasks for the simulator in the context of program analysis and verification. The first is to allow the user to explore program behaviour and read off details about its executions.
The other is to support verification tools which provide counterexamples to the user. As detailed in Section~\ref{sec:related}, it is a difficult task to analyse problem reports from automated
analysis and verification tools, and a simulator can be very helpful in this regard. In case of model checkers, the problem report contains an execution \emph{trace}: a step-by-step description of the
problematic behaviour. For tools based on \divm{}, this trace is simply a list of 2 types of information:

\begin{enumerate}
\def\labelenumi{\arabic{enumi}.}
\tightlist
\item
  The non-deterministic choices made during the execution of the program (internally, there is only one non-deterministic choice operator and all state-space branching is caused by this operator,
  including thread interleaving).
\item
  Which of the \emph{interrupt points} were used in the execution: the model checker may be able to prove that a particular interrupt point is not required, and the simulator needs this information to
  correctly reproduce the counterexample.
\end{enumerate}

Since the program is isolated from the environment, this list completely and unambiguously describes its entire execution history. When the model checker discovers a problem in the program, it writes
this list into a text file, which the simulator can then load along with the program.

When the simulator loads a trace, it locks the outcomes of all non-deterministic choices to follow the trace. In this mode, stepping through the program (backwards or forwards) will simply follow the
counterexample, unless a particular choice is overridden by the user. In effect, the user will be guided through the faulty behaviour of the program, and can easily move back and forth to locate the
cause of the problem (as opposed to the symptom, which is what the model checker reports and may be distinct from the original cause).

\hypertarget{sec:implementation}{%
\section{Implementation}\label{sec:implementation}}

The ideas presented in this paper are implemented in the simulator component of \divine{} 4, which is available as \texttt{divine\ sim}. All relevant source code is available online\footnote{\url{https://divine.fi.muni.cz/download.html}},
under a permissive open source licence. Additional details about the user interface and user interaction in particular can be found in the \divine{} 4 manual\footnote{\url{https://divine.fi.muni.cz/manual.html}}.

\hypertarget{user-interface}{%
\subsection{User Interface}\label{user-interface}}

The data structures and most of the code are independent of a particular user interface. In fact, two user interfaces exist for the simulator. The primary interface is command-driven, similar to
terminal-based symbolic debuggers like \texttt{gdb}. The command-line parser and other interface-specific code entails approximately 800 lines of C++. Additionally, a third-party graphical interface
is also available.\footnote{The source code of the graphical user interface is available from the supplementary materials page at \url{https://divine.fi.muni.cz/2019/sim/}.}

The command interface uses \emph{meta variables} extensively: each such meta variable holds a reference to a single debug node (cf. Section~\ref{sec:debugnode}). There are two basic types of meta
variables, \emph{static} and \emph{dynamic}.

Static variables always point to the same debug node, even as the program executes and the content of its memory changes. Since objects in the \divm{} memory are \emph{persistent} (not mutable), this
type of variable simply points to such a persistent, immutable object. Static meta variables have names starting with a \texttt{\#} sign, e.g. \texttt{\#start}.

Dynamic variables reflect the current state of the program at any given time. The debug nodes referenced by those variables are \emph{refreshed} every time the program mutates its memory, so that they
always point to an up-to-date copy of the persistent memory object (in other words, they always refer to the current program state). These variables are prefixed with a \texttt{\$} sign, e.g.
\texttt{\$frame}.

\hypertarget{programming-language-support}{%
\subsection{Programming Language Support}\label{programming-language-support}}

Our simulator design is, to a large degree, independent of the particular high-level language in which the simulated program was developed. The structure of the program is described in the debug info
metadata in sufficient detail to provide precise and readable information to the user. This is in contrast to tools like \texttt{gdb} and \texttt{lldb}~\citep{lee13:lldb} which mostly rely on
evaluating C and/or C++ statements for presenting the program data. That is, the user is allowed to type in a C or C++ expression to be evaluated and the result displayed. The major downside is that
if the high-level language support is incomplete (like it is the case with C++ support in \texttt{gdb}), it becomes much harder to obtain certain values without resorting to very low-level means
(printing bytes at particular addresses). Consequently, the amount of implementation work required to support a particular programming language in a debugger can be substantial.\footnote{We speculate
  that this is the primary reason why interactive simulators (and debuggers in general) are so scarce.}

On the other hand, the debug graph implemented in our simulator (see Section~\ref{sec:debugnode}) is language-neutral, and hence the features derived from this graph are independent of the original
programming language. For this reason, we consider the debug graph to be an important contribution: it can be built from \llvm{} debug metadata in a comparatively small amount of code, but nonetheless
provides a very convenient interface.

\hypertarget{sec:conclusion}{%
\section{Conclusion}\label{sec:conclusion}}

We have described a novel approach to interactive analysis of real, multi-threaded C and C++ programs. The approach plays an important support role in the wider context of automated verification and,
in particular, model checking of software. The simulator naturally supports the compact and universal counterexample format used in \divm{}. Compared to earlier tools, \divine{}~4 is substantially more
useful in practice, also thanks to the new interactive simulator.\footnote{Supported by anecdotal evidence from working with students, both individually and in a validation \& verification course.}

\bibliography{common}

\end{document}